\documentclass[preprint,5p,times]{elsarticle}
\usepackage{amssymb}
\usepackage{pstricks}
\usepackage{graphicx}
\usepackage{multirow}
\usepackage{amsmath}

\begin{document}

\begin{frontmatter}

\title{Low energy IceCube data and a possible Dark Matter related excess}

\author{M.~Chianese$^{a,b}$}
\author{G.~Miele$^{a,b}$}
\author{S.~Morisi$^{a,b}$}
\author{E.~Vitagliano$^{b,c}$}
\address[a]{INFN, Sezione di Napoli, Complesso Univ. Monte S. Angelo, I-80126 Napoli, Italy}
\address[b]{Dipartimento di Fisica {\it Ettore Pancini}, Universit\`a di Napoli Federico II, Complesso Univ. Monte S. Angelo, I-80126 Napoli, Italy}
\address[c]{Max-Planck-Institut f\"ur Physik (Werner-Heisenberg-Institut), F\"ohringer Ring 6, 80805 M\"unchen, Germany}

\begin{abstract}
In this Letter we focus our attention on the IceCube events in the energy range between 60 and 100 TeV, which show an order $2$-sigma excess with respect to a power-law with spectral index~2. We analyze the possible origin of such an excess by comparing the distribution of the arrival directions of IceCube events with the angular distributions of simply distributed astrophysical galactic/extragalactic sources, as well as with the expected flux coming from DM interactions (decay and annihilation) for different DM profiles. The statistical analysis performed seems to disfavor the correlation with the galactic plane, whereas rules out the DM annihilation scenario only in case of small clumpiness effect. The small statistics till now collected does not allow to scrutinize the cases of astrophysical isotropic distribution and DM decay scenarios. For this reason we perform a forecast analysis in order to stress the role of future Neutrino Telescopes.
\end{abstract}

\end{frontmatter}

\section{Introduction}

\begin{figure}[h!]
\begin{center}
\includegraphics[width=0.45\textwidth]{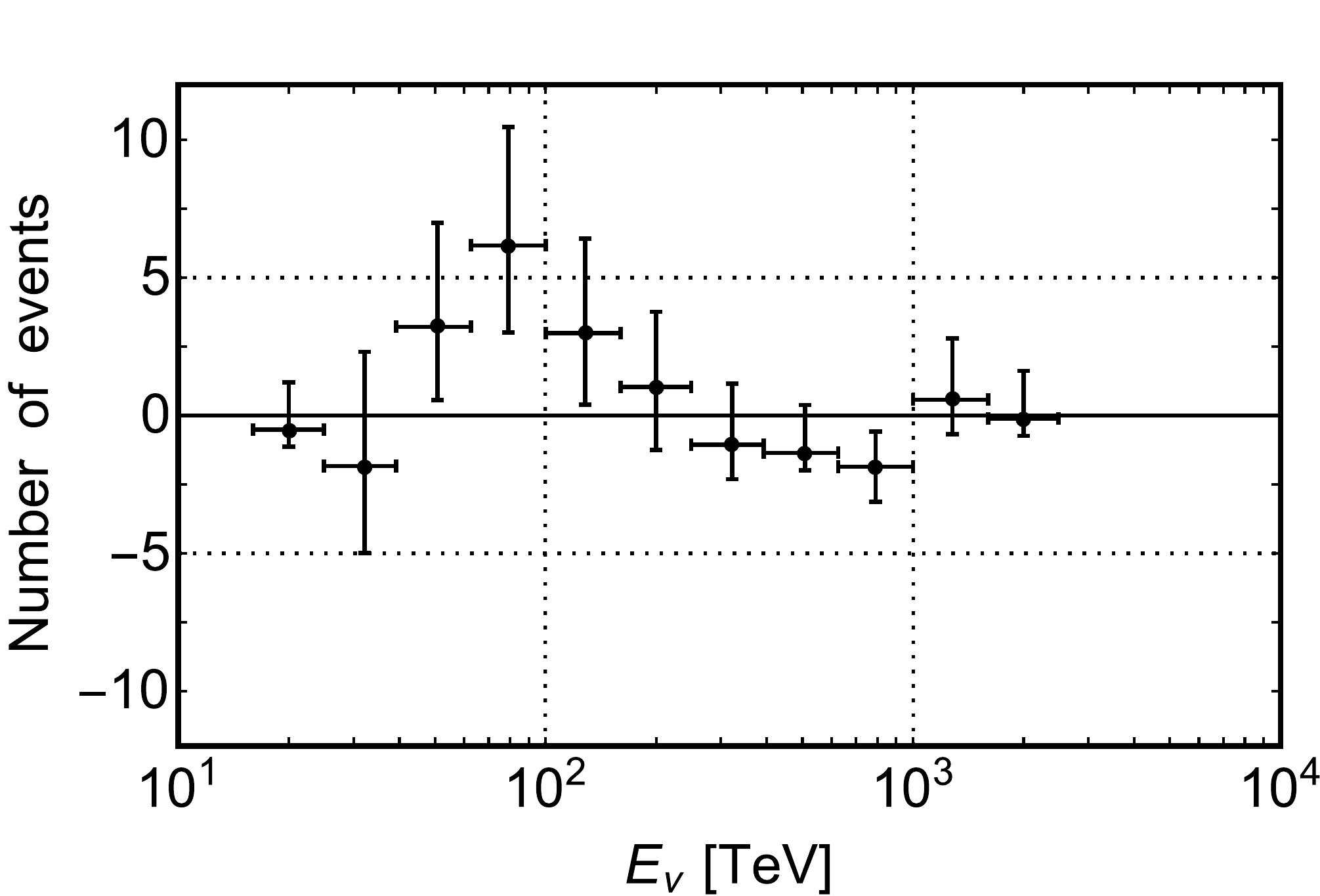}
\hskip2.mm
\includegraphics[width=0.45\textwidth]{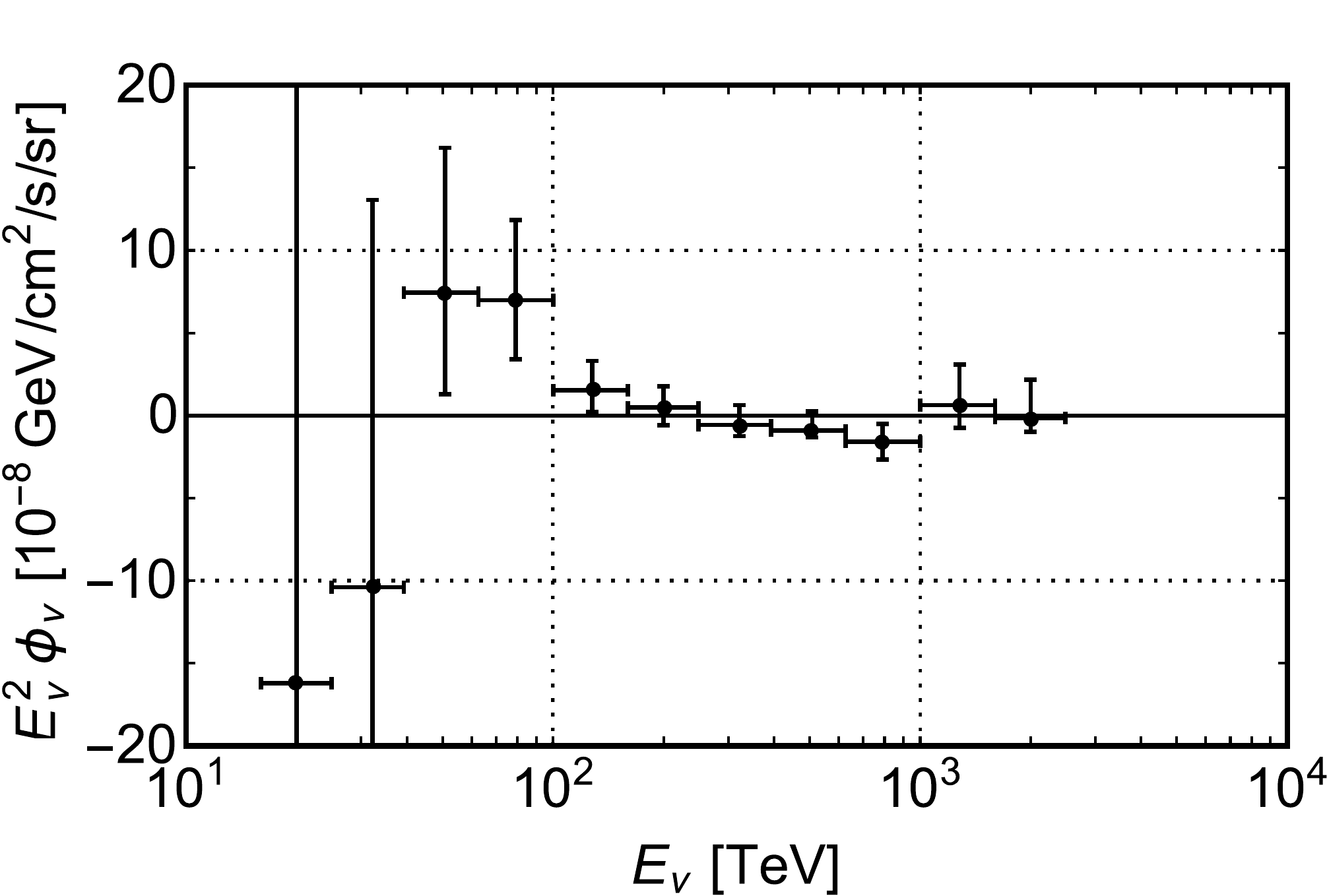}
\caption{The upper panel shows the excess in the number of IceCube neutrino events with respect to the sum of the background (atmospheric neutrinos and muons) and an astrophysical component described by a $E_\nu^{-2}$ power-law, as function of the neutrino energy. In the lower panel, we report the same excess in the whole neutrino flux (summed on all flavors) once that the average effective area of the particular energy bin and 1347 days of data taking have been taken into account.}
\label{fig1}
\end{center}
\end{figure}

The observation of astrophysical neutrinos made by the IceCube experiment (IC) \cite{Aartsen:2013jdh} has been an important step in the field of Neutrino Astronomy, whose impact on both Particle- and Astro-Physics has still to be unveiled. Recently \cite{Aartsen:2015zva} the IC Collaboration has delivered in a preliminary work the results of four years data. In Figure \ref{fig1} (upper panel) we report the so-called {\it excess} in the number of events as a function of the neutrino energy, which is the number of events detected in IC once one has subtracted the background (atmospheric conventional and prompt neutrinos plus muons  \cite{Aartsen:2015zva}), and an astrophysical component characterized by a $E_\nu^{-\gamma}$ power-law, with $\gamma=2$ being considered as benchmark prediction \cite{Aartsen:2013jdh,Aartsen:2015zva,Aartsen:2015knd}. The plot seems to suggest the presence of an excess of events in the energy range 60 - 100 TeV that has some tension, order $2\sigma$, with the simple $E_\nu^{-2}$ contribution. As can be seen from Ref.~\cite{Aartsen:2015zva}, such an excess partially disappears if one considers a steeper astrophysical component with power-law $E_\nu^{-2.58}$, even though such an exponent for a neutrino flux can be hardly explained.

In general, the cosmic ray spectrum is characterized by a power-law \cite{Gaisser:1990vg}, which is usually understood in terms of acceleration from shock fronts, the so-called Fermi mechanism \cite{Gaisser:1990vg,Bell:1978zc}. An astrophysical neutrino flux is expected to be produced during the hadronic matter acceleration and interaction with radiation ($p\gamma$) or with gas ($pp$). Thus its dependence on energy should be, at the source, mostly related to the differential spectrum of charged cosmic rays and to the pions production efficiency. If needed, one should take into account propagation effects so that, for example, the galactic cosmic ray spectrum at the Earth becomes $\propto E^{-(\gamma+\delta)}$ ($\gamma+\delta \approx 2.7$) up to the {\it ankle} at $1\sim10$~PeV \cite{Agashe:2014kda}. The quantity $\delta$ depends on galactic magnetic fields and is evaluated through cosmic rays secondary to primary ratio measurements \cite{Gaisser:1990vg,Swordy:1990yn}. On the other hand, galactic astrophysical neutrinos, which are not affected by magnetic fields, have a flux $\propto E_\nu^{-\gamma}$, with $\gamma\approx2$.

Even considering extragalactic sources to be significant (a reasonable hypothesis given the high galactic latitudes of some events) it is not easy to justify a steep flux \cite{Aartsen:2014gkd}. Models with $p\gamma$ interactions produce peaked spectra \cite{Mucke:1999yb}, so one could have a steep flux depending on the position of the peak. However, in this case one expects a peak in the spectra in the region of several PeV like for Active Galactic Nuclei \cite{Stecker:1991vm}.  Another kind of $p\gamma$ source, Gamma Ray Burst, provides a flux with an upper limit (given by searches for correlation with observed GRB) more than one order of magnitude below the observed flux \cite{Abbasi:2012zw}. Moreover, hard and smooth spectra are expected in $pp$ scenarios \cite{Kelner:2006tc}, like for extragalactic Supernova Remnants \cite{Chakraborty:2015sta}. More generally, theoretical models of acceleration mechanism for hadronic matter produce a flux that should be at most as soft as $E_\nu^{-2.2}$ \cite{Loeb:2006tw}\footnote{This is true when considering the minimal unbroken power law scenario; considering different values for $\delta$ \cite{Blasi:2011fm} or including a break in the neutrino spectrum can lead to a steep flux, which is however constrained by data (see below).}.

Independently of the detailed theoretical model, we have, most importantly, observational constraints coming from multimessenger approaches like \cite{Murase:2013rfa}, which combine IC data with Fermi $\gamma$-ray measurements, and give strong bounds to a hadronuclear origin (while leave room for hidden $p\gamma$ sources \cite{Murase:2015xka}). In fact, a spectrum as soft as the one obtained fitting IC data implies, when assuming sources transparent to radiation with respect to two-photon annihilation, an expected $\gamma$-ray flux bigger than the measured one \cite{Murase:2013rfa,Murase:2015xka}. 
The previous considerations support the assumption of an astrophysical $E_\nu^{-2}$ power-law used to obtain the excess reported in Figure \ref{fig1}.

Starting from the excess in the number of events one can obtain the corresponding quantity for the whole neutrino flux (summed on all flavors) once the {\it effective area} of the detector is taken into account \cite{Aartsen:2013jdh}. In Figure \ref{fig1} (lower panel) we report such a flux as a function of the neutrino energy. As already discussed in Ref.~\cite{Boucenna:2015tra}, we assume for simplicity the equality between the deposited and neutrino energy due  to low statistics at our disposal. At the energy scale $\mathcal{O}(100)$ TeV, this is not strictly true for neutral current interactions \cite{Aartsen:2013vja}. When a significant statistics is collected, the average ratio between the two energies, which is of the order of $(97\% \, \sigma^{\rm CC}+  23\%\, \sigma^{\rm NC})/(\sigma^{\rm CC} + \sigma^{\rm NC})\sim75\%$, could be applied.

In this Letter we assume that the above excess, mainly concentrated in the energy range 60 - 100 TeV, has a genuine physical origin. Under this ansatz, it is worth pursuing, for this energy bin, a study in order to unveil the nature of such an excess. We perform our analysis assuming as {\it null hypothesis} one of the following alternatives for the source of the IC data:\\ 
i) astrophysical, which can be investigated by studying, in first approximation, the correlation with the galactic plane or with an isotropic distribution for galactic or extragalactic astrophysical sources, respectively;\\
ii) induced by Dark Matter via decay or annihilation, hence related to the first or second power of the particular Dark Matter (DM) density profile adopted.\\
Moreover, even though the small number of events already detected does not allow to exclude all DM scenarios, one can perform a forecast analysis in order to determine the required statistics.

In order to compare the IC observations with possible DM predictions we consider both decaying and stable Dark Matter cases. In the first case, $60-100$~TeV neutrinos detected at IceCube would be originated directly from the decay of the DM particles, while for a stable DM particle neutrinos are only produced {\it via} annihilation. In both cases the resulting neutrino flux would be composed by both a galactic and an extragalactic DM component. Different approaches proposed in literature \cite{Boucenna:2015tra,Feldstein:2013kka,Barger:2013pla,Esmaili:2013gha,Bai:2013nga,Ema:2013nda,Bhattacharya:2014vwa,Higaki:2014dwa,Ema:2014ufa,Rott:2014kfa,Esmaili:2014rma,Fong:2014bsa,Dudas:2014bca,Murase:2015gea,Ko:2015nma,Aisati:2015ova} have studied the possible presence of DM hints in the PeV range, namely for the most energetic IC events. Here we take an alternative point of view, assuming that PeV events have bottom-up origin and considering for lower energy data a possible top-down origin due to  DM particles with mass scale $\mathcal{O}(100)$~TeV. Independently of the mass scale and of the DM couplings, neutrinos originated from DM would have an angular distribution that is more peaked around the Galactic Center  where a higher DM density is expected. This is true in particular when assuming an annihilating DM, because of the squared enhancement factor. Of course, this effect is dependent on the assumed DM galactic halo profiles; for example, one could take the Navarro-Frenk-White profile (NFW) \cite{Navarro:1995iw} or different distributions like the Isothermal profile (Isoth.), which implies a more isotropic flux.

\section{The analysis}

\begin{figure}[h!]
\begin{center}
\includegraphics[width=0.45\textwidth]{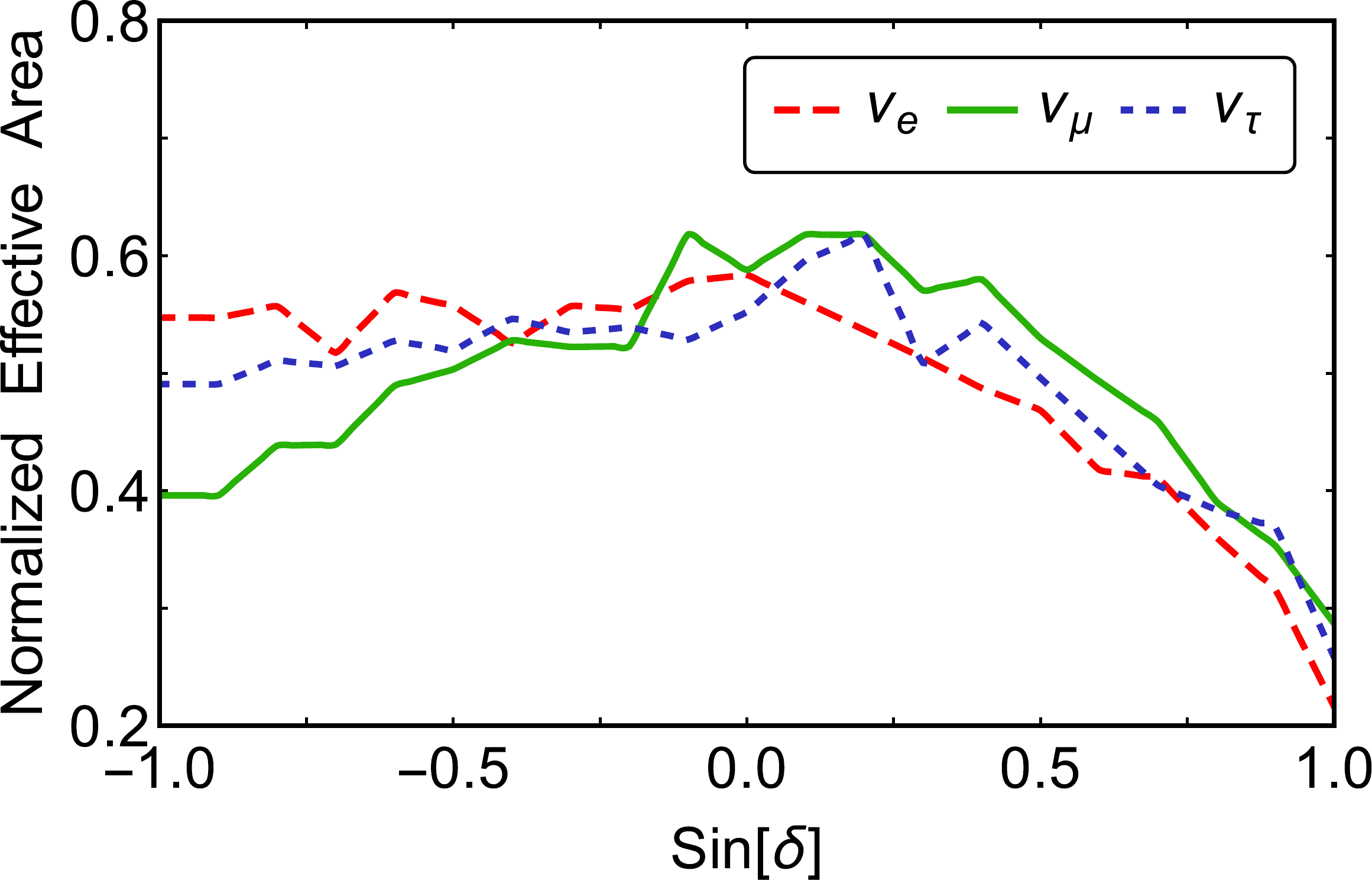}
\caption{Normalized IceCube effective area as function of the declination $\delta$, once it has been averaged in the energy range $60-100$\,TeV.}
\label{fig2}
\end{center}
\end{figure}

In order  to infer about the physical origin of the excess we compare the angular distribution of the observed events in the energy bin 60 - 100 TeV (in the following we discuss the procedure to take into account the presence of the background and of the experimental errors) with the angular distributions of astrophysical galactic sources (galactic plane) and extragalactic ones (isotropic distribution), as well as with the expected flux coming from DM interactions (decay and annihilation). In this approach the astrophysical $E^{-2}$ power-law contribution is regarded just as an additional term to the background events counting for atmospheric neutrinos and muons. For this reason hereafter we denote  as {\it background} the sum of atmospheric neutrinos, muons and neutrinos coming from the astrophysical $E^{-2}$ power-law. Moreover,  due to the small number of events collected till now in the energy bin under study, in this analysis we take the simplicity assumption to consider just one additional component to neutrino background at a time (alternative scenarios i) or ii) of previous section) to explain the excess. This allows us to be more predictive even though more involved scenarios can be proposed where the excess in neutrino flux can be explained in terms of several components of different origin. Other analyses have already been presented in literature with different assumptions \cite{Esmaili:2014rma,Troitsky:2015cnk}. In particular, in Ref. \cite{Esmaili:2014rma} it has been studied the possibility that the whole neutrino spectrum has a DM origin, while Ref. \cite{Troitsky:2015cnk} analyzed only the high energy neutrino spectrum ($E_\nu>150$ TeV) considering also mixed components to the neutrino flux. 
Differently from previous analyses \cite{Esmaili:2014rma,Troitsky:2015cnk}, we also take into account the angular efficiency of the IC detector for all neutrino flavors \cite{icecube_data}. In Figure \ref{fig2} it is reported the normalized IC effective area, averaged in the energy range considered ($60-100$\,TeV), as function of $\sin\delta$, where $\delta$ is the declination in the equatorial coordinates system. In the region $\sin\delta>0$ (North Pole) the effective area is mainly affected by the Earth absorption for all neutrino flavors, whereas for $\sin\delta<0$ (South Pole) only the muon effective area is slightly dependent on the declination. 

In order to simplify the description, in the following we obtain the expected neutrino angular distributions for the analyzed cases without explicitly considering the IC effective area, which will be instead considered  
for the real analysis.

In case of astrophysical scenarios, namely {\it galactic plane} and {\it isotropic distribution}, the expected angular distributions in the arrival directions are
{\small{
\begin{eqnarray}
 p^{\rm{gal}}(\sin b,l)&=&\frac{\Theta(\sin b+\sin b_{\rm gal})-\Theta(\sin b-\sin b_{\rm gal})}{4 \pi  \sin b_{\rm gal}}\,,\\
 p^{\rm{iso}}(\sin b,l)&=&\frac{1}{4 \pi}\,.
\end{eqnarray}}}
Note that the galactic plane angular distribution depends on the Galactic latitude $b$ only ($l$ being the Galactic longitude), where $b_{\rm gal}$ represents the angular size of the Galactic disk. It is reasonable to assume that the galactic neutrino angular distribution has the same characteristics of the galactic gamma-ray one \cite{Neronov:2014uma}. Using the Fermi-LAT template \cite{Ackermann:2012pya}, the quantity $b_{\rm gal}$ has been varied in the range $\left[2^{\circ},4^{\circ}\right]$. We assume for simplicity that the astrophysical sources are uniformly distributed in the Galactic disk. Indeed, it is worth observing that, however, such an approximation does not dramatically change the qualitative result of our analysis, which is mainly affected by the low statistics at our disposal.

In case of decaying DM scenario, the expected differential neutrino flux results to be a sum of the contribution due to the {\it Galactic} halo, and of an {\it Extra Galactic} term
\begin{equation}
\frac{{\rm d}J}{{\rm d}\Omega {\rm d}E}  =  \frac{{\rm d}J^{\rm G}}{{\rm d}\Omega {\rm d}E} + \frac{{\rm d}J^{\rm EG}}{{\rm d}\Omega {\rm d}E}\label{flux0}
\end{equation}
where
\begin{eqnarray}
\frac{{\rm d}J^{\rm G}}{{\rm d}\Omega {\rm d}E}  & = & \frac{f\left(E\right)}{4 \pi M \tau}  \int_0^\infty\rho_h[r(s,\cos\theta)] {\rm d}s \, ,\label{fluxG}\\
\frac{{\rm d}J^{\rm EG}}{{\rm d}\Omega {\rm d}E}  & = & \frac{\Omega_{\rm DM}\rho_c}{4 \pi M \tau}  \int_0^\infty \frac{{\rm d}z}{H(z)}\left.f\left(E'\right)\right|_{E'=E(1+z)}.
\label{fluxEG}
\end{eqnarray} 
In the previous expressions $M$ and $\tau$ stand for mass and lifetime of DM particles\footnote{These parameters define an overall constant that results irrelevant in the present analysis.}, whereas  $\rho_h (r)$ is the DM galactic halo profile, being $r(s,\cos\theta)=\sqrt{s^2+R^2-2sR\cos{\theta}}$ with $R\simeq8.5\ \rm{kpc}$ (the Sun-Galactic Center distance) and $\cos\theta \equiv \cos b \cos l$. Moreover we take $\Omega_{\rm DM}=0.2685$, $\rho_c=5.5\times10^{-6}\rm{GeV cm}^{-3}$ and $H(z)=H_0\sqrt{\Omega_\Lambda+\Omega_m(1+z)^3}$ with $h\equiv H_0/100\,{\rm km}\,{\rm s}^{-1}\,{\rm Mpc}^{-1}$, $\Omega_\Lambda=0.6825$ and $\Omega_m=0.3175$ \cite{Ade:2015xua}. The quantity $f\left(E\right)\equiv{{\rm d}N}/{{\rm d}E}$ is the neutrino energy spectrum produced by the decay of a DM particle, whose expression depends on the specific model considered. Since the excess is localized in the energy range 60 - 100 TeV, it is reasonable to assume that the neutrino energy spectrum is almost peaked in the same range. This means  to assume a negligible neutrino energy spectrum for energy larger than 100~TeV. Note that a not vanishing neutrino energy spectrum for energy smaller that 60~TeV does not provide any contribution neither to the galactic component nor to the extragalactic one for the energy bin considered. Therefore, independently of the shape of the energy distribution $f\left(E\right)$, the redshift integral in Eq.~(\ref{fluxEG}) has an upper limit that is at most equal to $z_{\rm{max}}=100~{\rm{TeV}}/60\,{\rm{TeV}}-1$. This is determined  by the relation $E^\prime=E\left(1+z\right)$  when  the conditions $E^\prime\leq 100\,{\rm{TeV}}$ and $60\,{\rm{TeV}}\leq E \leq 100\,{\rm{TeV}}$ are applied. Therefore, by integrating Eq.~(\ref{fluxEG}) in the 60 - 100 TeV range one obtains
{\small{
\begin{eqnarray}
&&\frac{{\rm d}J^{\rm EG}}{{\rm d}\Omega} \simeq  \frac{\Omega_{\rm DM}\rho_c}{4 \pi M \tau} \int_0^{z_{\rm max}} \frac{{\rm d}z}{(1+z)H(z)} \nonumber \\
& & \times\left\{ \int_{E_{\rm m}}^{E_{\rm M}}f\left(E\right){\rm d}E  + \left[ f\left(E_{\rm M}\right) E_{\rm M}-  f\left(E_{\rm m}\right)E_{\rm m}\right]\,z\right\}\, ,
\label{flux_EG}
\end{eqnarray}}}
where $E_{\rm m}=60~{\rm TeV}$ and $E_{\rm M}=100~{\rm TeV}$. There exist two interesting limits:\\ {\it i)} the neutrino energy spectrum is fully contained in the energy bin considered, then $f\left(E_{\rm m}\right) \simeq f\left(E_{\rm M}\right) \simeq 0$; \\{\it ii)}  the neutrino energy spectrum is wider than the energy bin, hence the energy distribution $f\left(E\right)$ can be considered almost flat within it, namely $f\left(E_{\rm m}\right) \simeq f\left(E_{\rm M}\right)$.\\ This implies that the angular distribution takes the form
\begin{equation}
 p^{\rm{dec}}(\cos\theta) \propto \int_0^\infty\rho_h[r(s,\cos\theta)]{\rm d}s + \Omega_{\rm DM}\rho_c\,\beta_\alpha \,,
\end{equation}
with
\begin{equation}
\beta_\alpha=\int_0^{z_{\rm{max}}} \frac{{\rm d}z}{(1+z)^\alpha H(z)}\,,
\end{equation}
where $\alpha=1$ and $0$ in the first and second case respectively, so providing  $\beta_1=0.43/H_0$ and $\beta_0=0.56/H_0$. In the following, to be more conservative we consider only the case corresponding to a larger extragalactic contribution ($\alpha=0$), since in this case the isotropic cosmological contribution results to be more competitive with the Galactic term. It is worth observing that including explicitly  the IC effective area in Eq.~(\ref{flux_EG}) does not change this result, and the case with $\alpha=0$ still represents the most conservative scenario.\\ Similarly, one obtains the following angular distribution in case of annihilating DM scenario
\begin{equation}
p^{\rm{ann}} (\cos\theta) \propto \int_0^\infty\rho^2_h[r(s,\cos\theta)]{\rm d}s + (\Omega_{\rm DM}\rho_c)^2\,\Delta^2_0 \,\beta_\alpha\,,
\end{equation}
where $\Delta^2_0$ is the clumpiness factor \cite{Hooper:2007be} (see also Ref.~\cite{Cirelli:2010xx}). The quantity $\Delta^2_0$ ranges from $10^4$ to $10^8$ \cite{Taylor:2002zd} depending on the model considered. In our analysis we consider three particular cases where $\Delta^2_0$ is equal to $10^4$, $10^6$ and $10^8$ corresponding to an extragalactic contribution that is sub-dominant, comparable and dominant with respect to the galactic one, respectively. However, recent studies like \cite{Sefusatti:2014vha} state that the cumpliness factor $\Delta^2_0$ can be as large as few times $10^6$, considering unphysical larger values for such a quantity.

In our analysis, we consider two different DM galactic halo profiles \cite{Cirelli:2010xx}: the Navarro-Frenk-White distribution
\begin{equation}
\rho_h^{\rm{NFW}}\simeq\frac{\rho_h}{r/r_c(1+r/r_c)^2} \, ,
\end{equation}
where $r_c\simeq 20\ \rm{ kpc}$ and $\rho_h=0.33\ \rm{ GeV cm}^{-3}$, and the Isothermal distribution
\begin{equation}
\rho_h^{\rm{Isoth}}\simeq\frac{\rho_h}{1+(r/r_c)^2}\,,
\end{equation}
where  $r_c\simeq 4.38\ \rm{kpc}$ and $\rho_h=1.39\ \rm{ GeV cm}^{-3}$.

Since in each case the distributions depend on one angle only, we can perform a {\it one-dimensional} statistical test. In particular, we use two different {\it non-parametric} statistical tests: the Kolmogorov-Smirnov test (KS) \cite{kstest} and the Anderson-Darling test (AD) \cite{adtest}. These statistical tests make a comparison between the cumulative distribution function (CDF) of the null hypothesis distribution function and the empirical cumulative distribution function (EDF), given by
\begin{equation}
{\rm EDF}(\cos\theta)=\frac{1}{n} \sum_{i=1}^n \Theta \left(\cos\theta - \cos\theta_i \right)\,,
\end{equation}
where $n$ is the number of observed events $\cos\theta_i$. Note that, in case of galactic plane angular distribution, the variable $\cos\theta$ has to be changed into $\sin b$. In the Kolmogorov-Smirnov test, the Test Statistics (TS) is the maximum distance between the previous two cumulative distribution functions and it is defined as
\begin{equation}
{\rm TS_{KS}} \equiv  {\sup}_{\theta}\left|{\rm EDF}(\cos\theta) - {\rm CDF}(\cos\theta) \right|\,,
\end{equation}
whereas in the Anderson-Darling test the Test Statistics is given by
\begin{eqnarray}
{\rm TS_{AD}} & \equiv & -n-\frac{1}{n}\sum_{i=1}^n (2i-1)\left[\ln \left( {\rm CDF}(\cos\theta_i)\right)\right. \nonumber\\
&& \left.+ \ln \left( 1 - {\rm CDF}(\cos\theta_{n+1-i}\right) \right] \,.
\label{eq:ts_ad}
\end{eqnarray}
In particular, this expression is very sensitive to the difference between the functions EDF and CDF at the two endpoints, suggesting that the Anderson-Darling test is a suitable test for our analysis (note that the Galactic Center is in correspondence of $\cos\theta = 1$).

To take into account the atmospheric background, we consider all possible different choices of 5 background events among 12, namely $12! /(5! \, 7!) = 792$ combinations. Moreover, we include in our analysis the angular uncertainty affecting the reconstruction of the arrival direction for IC events, which for the shower-like topology is very large, namely of the order of 15$^\circ$. In particular, we treat the uncertainties on declination and right ascension as maximum errors, and propagate them on the quantity $\cos\theta$. Note that for galactic plane scenario the variable to be considered is the Galactic latitude $b$.

To consider in our statistical tests the above angular uncertainty, for each choice of 5 background events, we consider 100 possible extractions of the 7 remaining events from their maximum error intervals using a uniform probability. In this way, for the 100 different choices of observed events we compute the corresponding TS values, which once compared with the null hypothesis TS distribution, provide a range of p-values. Such a range is finally averaged on the 792 different background combinations. In Table~\ref{tab1} we report such an average range for each test.
\begin{table}[h!]
\begin{center}
\begin{tabular}{|c|c|c|c|}
\hline
\multicolumn{2}{|c|}{{\bf Scenario}} & {{\bf KS}} & {{\bf AD}} \\ \hline\hline
\multirow{2}{*}{{{\small Astrophysics}}} & {\small Gal. plane} & {\small0.007 - 0.008} & {\small not defined} \\ \cline{2-4}
& {\small Iso. dist.} & {\small0.20 - 0.55} & {\small0.17 - 0.54} \\ \hline \hline
\multirow{2}{*}{{{\small DM decay}}} & {\small NFW} & {\small0.06 - 0.16} & {\small0.03 - 0.14} \\ \cline{2-4}
& {\small Isoth.}& {\small0.08 - 0.22} & {\small 0.05 - 0.19}\\ \hline \hline
{\small DM annih.} & {\small NFW} & {\footnotesize $\left(0.3 - 0.9\right)\times 10^{-4}$} & {\footnotesize $\left(0.3 - 3.8\right)\times 10^{-4}$} \\ \cline{2-4}
{\small $\Delta^2_0=10^4$} & {\small Isoth.} & {\footnotesize$\left(0.9 - 2.8\right)\times 10^{-3}$} & {\footnotesize$\left(1.0 - 5.0\right)\times 10^{-3}$}\\ \hline
{\small DM annih.} & {\small NFW} & {\small0.02 - 0.05} & {\small0.02 - 0.07} \\ \cline{2-4}
{\small $\Delta^2_0=10^6$} & {\small Isoth.} & {\small0.10 - 0.28} & {\small0.08 - 0.29} \\ \hline
{\small DM annih.} & {\small NFW} & {\small0.19 - 0.54} & {\small0.17 - 0.53}\\ \cline{2-4}
{\small $\Delta^2_0=10^8$} & {\small Isoth.} & {\small0.20 - 0.55} & {\small0.17 - 0.54}\\ \hline
\end{tabular}
\caption{Background average range of $p$-values for all the scenarios, using the Kolmogorov-Smirnov and the Anderson-Darling tests.}
\label{tab1}
\end{center}
\end{table}
As we can see from Table~\ref{tab1}, the IC data indicate that a correlation with the galactic plane is disfavored. Note that in this case, the Anderson-Darling test is not well defined since its CDF is vanishing within the region $b<b_{\rm gal}$ (see Eq. (\ref{eq:ts_ad})). It is worth observing that varying the angular size $b_{\rm gal}$ in the range $\left[2^{\circ},4^{\circ}\right]$ does not significantly change the $p$-value range reported in the Table. Moreover, the DM annihilation scenario is already excluded from IC data for both DM halo density profiles in case of a small clumpiness factor $(\Delta_0^2=10^4)$. On the other hand, for a larger clumpiness factor $(\Delta_0^2=10^8)$ we get a result similiar to the one of the astrophysical isotropic distribution. This is due to the fact that in this case the annihilating DM angular distribution is almost isotropic. It is worth observing that due to a certain lack of events from the Galactic Center, the NFW DM profile that is more peaked in this central region results to be more in tension with the observations than the Isothermal profile. This results in smaller p-values for NFW with respect to Isothermal as shown in the Table, such difference is exacerbated for annihilating DM scenario.

\section{Forecast}

It is of interest to ask about the statistics required (number of events) in order to distinguish, at a certain confidence level, a DM induced distribution from an isotropic one. To answer this question we perform a {\it forecast} analysis restricted to decaying DM scenario and annihilating DM one with $\Delta_0^2=10^6$ that are not already excluded by present data. For a given number of events, we generate $10^5$ sets of data (in the 60 - 100~TeV energy range) according to the isotropic distribution, and perform the two statistical tests under null hypothesis that the data samples come from a decaying DM  distribution or from an annihilating DM one. For simplicity we assume that each data sample is not affected by the background. To include the background effect in the forecast analysis one can simply increase our ``predictions'' by a factor of  $\sim 12/7$ as suggested by present data.

\begin{figure}[h!]
\begin{center}
\includegraphics[width=0.48\textwidth]{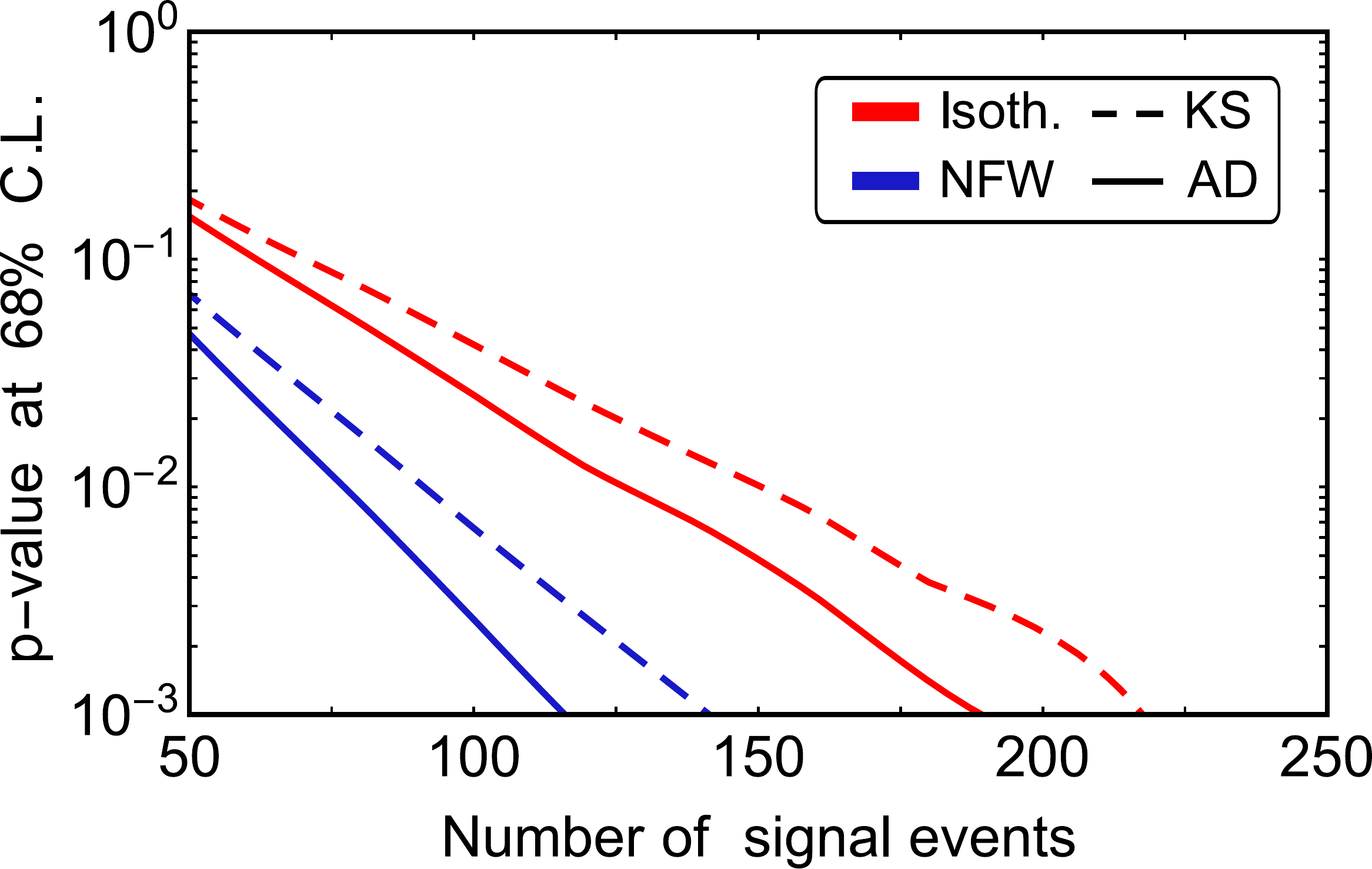}
\caption{Forecast analysis in case of decaying DM scenario for NFW (blue, lower) and Isothermal (red, upper) halo density profiles. The solid (dashed) lines are related to the Anderson-Darling (Kolmogorov-Smirnov) statistical test.}
\label{fig3}
\end{center}
\end{figure}

By varying in the set of $10^5$ data samples we get a distribution of p-value for which it can be defined  the $p$-value at 68\% Confidence Level (C.L.). This value represents the upper bound for p-values in 68$\%$ of cases. In Figure~\ref{fig3} we report the $p$-value at 68\% C.L. as function of the number of signal events (no background) in case of decaying DM scenario.
\begin{figure}[h!]
\begin{center}
\includegraphics[width=0.48\textwidth]{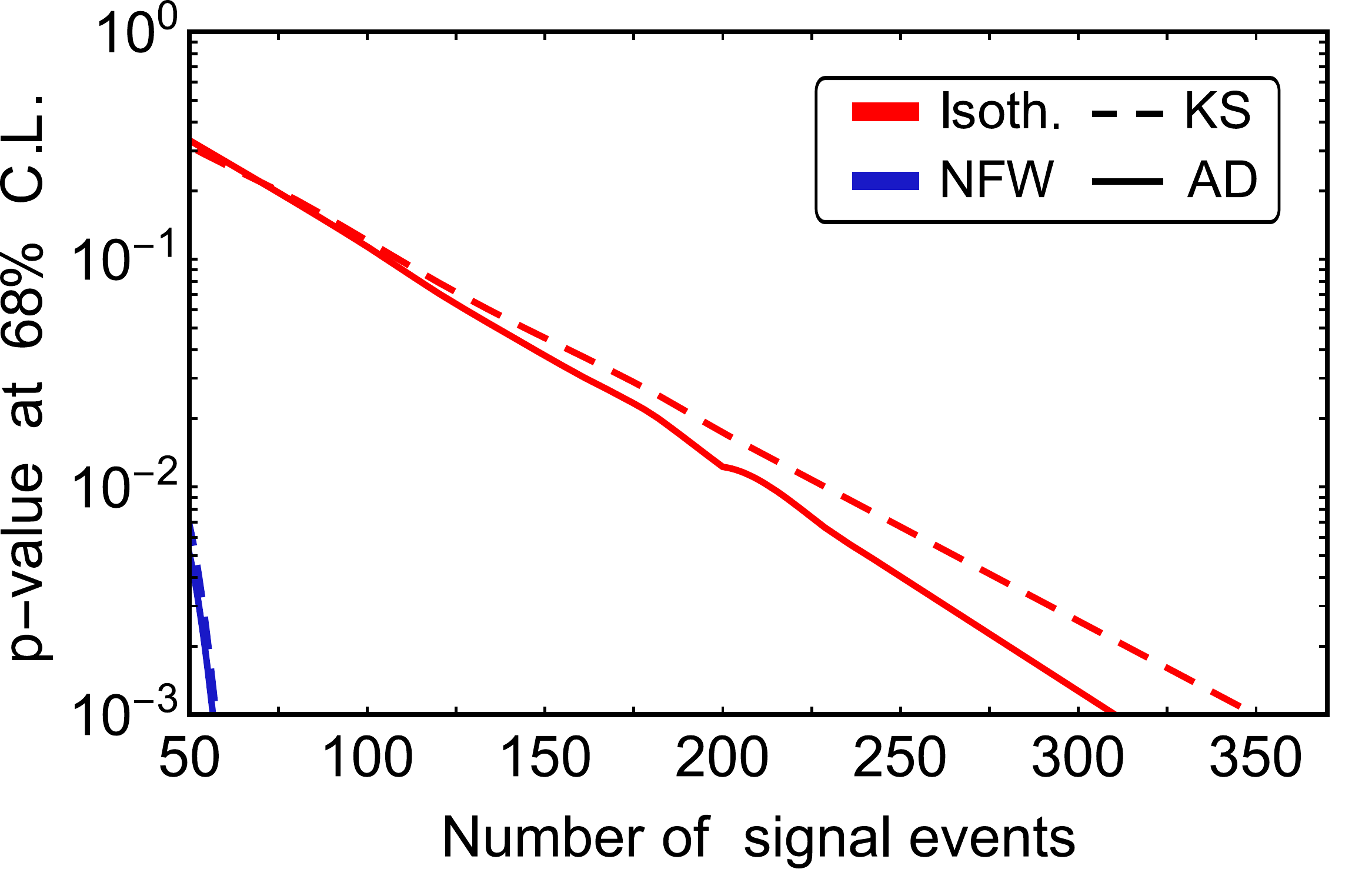}
\caption{Forecast analysis in case of annihilating DM scenario with clumpiness factor $\Delta^2_0=10^6$ for NFW (blue, lower) and Isothermal (red, upper) halo density profiles. The solid (dashed) lines are related to the Anderson-Darling (Kolmogorov-Smirnov) statistical test.}
\label{fig4}
\end{center}
\end{figure}
As expected, the Anderson-Darling statistical test (solid lines) is more appropriate than the Kolmogorov-Smirnov one (dashed lines). Indeed, the p-value falls down to zero very rapidly. Assuming that the p-value required to exclude a model is $\mathcal{O}(10^{-3})$, we see that the decaying DM scenario will be completely excluded only when a $\mathcal{O}(200)$ number of signal events is collected in the energy bin 60 - 100 TeV.  It is worth noticing that the NFW density profile, since more spatially concentrated around the Galactic Center, requires a small number of signal events, namely $\mathcal{O}(100)$, to be excluded with respect to the Isothermal profile. The forecast analysis in case of annihilating DM scenario with intermediate value of clumpiness $\left( \Delta^2_0=10^6\right)$ is shown in Figure~\ref{fig4}. In particular, we have obtained that in order to exclude such a scenario the required number of signal events is $\mathcal{O}(300)$. Such a huge statistics, even though cannot be reached in the present experimental set up, could be eventually reached in future Neutrino Telescopes \cite{Aartsen:2014njl,km3net}.

\section {Conclusions and outlook}
 
In this Letter we have analyzed the $60-100$~TeV bin of IceCube data that seems to suggest the presence of a $\sim 2 \sigma$ excess once the sum of the background (atmospheric neutrinos and muons) and an astrophysical component described by a $E_\nu^{-2}$ power-law is subtracted. In order to get information on the possible origin of such an excess, we have compared the distribution of the arrival directions of IceCube data with the angular distributions of simply distributed astrophysical galactic/extragalactic sources (galactic plane/isotropic distribution), as well as with the expected flux coming from DM interactions (decay and annihilation) for different DM profiles. The statistical analysis performed by means of Kolmogorov-Smirnov  and Anderson-Darling tests of hypothesis seems to disfavor the correlation with the galactic plane, whereas excludes the annihilating DM scenario for both NFW and Isothermal density profiles in case of a small clumpiness factor $(\Delta_0^2=10^4)$. The small number of events till now collected does not allow to distinguish between the case of an isotropic distribution (astrophysical extragalactic sources and annihilating DM scenario with a large clumpiness factor) and the remaining cases (DM decay scenarios and DM annihilation ones with $\Delta_0^2=10^6$). In this concern we have performed a forecast analysis and we have found that $\mathcal{O}(200)$ ($\mathcal{O}(300)$) signal events are required in order to exclude the decaying DM scenario (annihilating DM scenario with $\Delta^2_0=10^6$). If the lack of neutrino events towards the Galactic Center is confirmed by future data, the NFW decaying DM scenario will be more easily excluded.

\section*{\bf Acknowledgments}

We thank the IceCube collaboration for providing us the effective areas and in particular Claudio Kopper, and we thank Elisa Bernardini and Pasquale Dario Serpico for useful discussions and Sergio Palomares Ruiz. The authors acknowledge support by the Instituto Nazionale di Fisica Nucleare I.S. TASP and the PRIN 2012 ``Theoretical Astroparticle Physics" of the Italian Ministero dell'Istruzione, Universit\`a e Ricerca.

\end{document}